\documentclass[a4paper,12pt]{article}
\usepackage{amsmath}
\usepackage{amssymb}

\begin{document}

\title{Branching Ratios of $B  \to D_s K$ Decays \\
in Perturbative QCD Approach}

\author{Cai-Dian L\"u$^{a,b}$\footnote{e-mail: lucd@ihep.ac.cn}
  and Kazumasa Ukai$^c$\footnote{e-mail: ukai@eken.phys.nagoya-u.ac.jp}\\
   {\it \small $a$ CCAST (World Laboratory), P.O. Box 8730,
   Beijing 100080, China}\\
{\it \small $b$ Theory Division, Institute of High Energy Physics,
CAS, P.O.Box 918(4) }\\
{\it \small Beijing 100039, China}\footnote{Mailing address.}\\
{\it \small $c$ Physics Department, Nagoya University, Nagoya 464-8602, Japan}}

\maketitle

\begin{picture}(0,0)(-320,-320)
 \put(0,32){\bf hep-ph/0210206}
 \put(0,15){BIHEP-TH-2002-47}
 \put(0,0){DPNU-02-31}
\end{picture}

\begin{abstract}
The rare decays $B^0  \to D_s^- K^+$
and $B^+ \to D_s^+ \overline{K}^0$ can occur only
via annihilation type diagrams in the standard model.
We calculate these decays
in  perturbative QCD approach.
We found that the calculated branching ratio of  $B^0  \to D_s^- K^+$
agreed with the data which had been observed in the KEK and SLAC $B$
 factories.
While the decay $B^+ \to D_s^+ \overline{K}^0$ has a very small branching ratio
at ${\cal O}( 10^{-8})$, due to the suppression from
CKM matrix elements $|V_{ub}^* V_{cd}|$.
\end{abstract}

\newpage
\section{Introduction}

The rare $B$ decays are being measured at $B$ factories in KEK and SLAC.
The generalized factorization approach has been applied to the
theoretical treatment of non-leptonic  $B$ decays \cite{bsw}.
It is a great success in explaining many decay branching ratios
\cite{akl1,cheng}.
The factorization approach (F.A.) is a rather simple method.
Some efforts have been made to improve their theoretical application
\cite{bbns} and to understand the reason why the F.A.
has gone well \cite{Keum:2000,Lu:2000em}.
One of these method is the perturbative QCD approach (PQCD),
where we can calculate the annihilation diagrams as well as the factorizable
and nonfactorizable diagrams.

The rare decays $B \to D_s K$ are pure annihilation type decays.
In the usual F.A., this decay picture is described as
$B$ meson annihilating into vacuum and the $D_s$ and $K$ mesons produced
from vacuum then afterwards.
To calculate this decay in the F.A., one needs the
$D_s \to K$ form factor at very large time like momentum transfer
${\cal O} (M_B)$.
However the form factor at such a large momentum transfer is not known
in F.A.
This makes the F.A. calculation of these decays unreliable.
The annihilation amplitude is a phenomenological parameter in QCD
factorization approach (QCDF)\cite{bbns}, and the QCDF calculation of
these decays is also unreliable.
In this paper, we will try to use the PQCD approach,
where the annihilation amplitude is calculable,
to evaluate the $B \to D_s K$ decays.
By comparing the predictions with the experimental data,
we can test the PQCD evaluation of the annihilation amplitude.

A $W$ boson exchange causes $\bar{b}d \to \bar{c}u $ or
$\bar b u \to \bar{d}c$, and
the $\bar{s}s$ quarks included in $D_s K$ are produced from a gluon.
This gluon  attaches to any one of the quarks participating in the $W$
boson exchange.
This is shown in Figure \ref{fig:need1gluon}.
In the rest frame of $B$ meson,
both $s$ and $\bar{s}$ quarks included in $D_s K$ have
$\mathcal{O}(M_B/2)$ momenta, and the gluon producing them also has
$q^2 \sim \mathcal{O}(M_B^2/4)$. This is a hard gluon.
One can perturbatively treat the process where the four quark operator
 exchanges a hard gluon with $s \bar s$ quark pair.
It is just the picture of PQCD approach.

In the next section, we explain the framework of PQCD briefly.
In section \ref{sc:formula},
we give the analytic formulas for the decay amplitude
of $B  \to D_s K$ decays.
In section \ref{sc:neval}, we show the predicted branching ratio from the
analytic formulas and discuss the theoretical errors.
It is found that the prediction are is in good agreement
with the data, and PQCD approach correctly gives the annihilation amplitude
in consequence.
Finally, we conclude this study in section \ref{sc:concl}.

\section{Framework}\label{sc:fm}

 PQCD approach has been developed and applied in the non-leptonic $B$
meson decays
\cite{Chang:1997dw,Keum:2000,Lu:2000em,PQCD}
for some time.
In this approach,
the decay amplitude is separated into
soft($\Phi$), hard($H$), and harder($C$) dynamics characterized by
different scales.
It is conceptually written as the convolution,
\begin{equation}
 \mbox{Amplitude}
\sim \int\!\! d^4k_1 d^4k_2 d^4k_3\
\mathrm{Tr} \bigl[ C(t) \Phi_B(k_1) \Phi_{D_s}(k_2) \Phi_K(k_3)
H(k_1,k_2,k_3, t) \bigr],
\label{eq:convolution1}
\end{equation}
where $k_i$'s are momenta of light quarks included in each mesons, and
$\mathrm{Tr}$ denotes the trace over Dirac and color indices.
$C(t)$ is Wilson coefficient which results from the radiative
corrections at short distance.
In the above convolution, $C(t)$ includes the harder dynamics
at larger scale than $M_B$ scale and describes the evolution of local
$4$-Fermi operators from $m_W$, $W$ boson mass, down to
$t\sim\mathcal{O}(\sqrt{\bar{\Lambda} M_B})$ scale, where
$\bar{\Lambda}\equiv M_B -m_b$.
$H$ describes the four quark operator and the spectator quark connected by
 a hard gluon whose $q^2$ is on the order
of $\bar{\Lambda} M_B$, and includes the
$\mathcal{O}(\sqrt{\bar{\Lambda} M_B})$ hard dynamics.
Therefore, this hard part $H$ can be perturbatively calculated.
$\Phi_M$ is the wave function which describes hadronization of the quark
and anti-quark to the meson $M$.
While the $H$ depends on the processes considered,
$\Phi_M$ is  independent of the specific processes.
Determining $\Phi_M$ in some other decays,
we can make quantitative predictions here.

We consider the $B$ meson at rest for simplicity.
It is convenient to use light-cone coordinate $(p^+, p^-, {\bf p}_T)$
to describe the meson's momenta, where 
$p^\pm = \frac{1}{\sqrt{2}} (p^0 \pm p^3)$ and ${\bf p}_T = (p^1, p^2)$.
Using these coordinates we can take the $B$, $D_s$, and $K$ mesons momenta
as
$P_1 = \frac{M_B}{\sqrt{2}} (1,1,{\bf 0}_T)$,
$P_2 = \frac{M_B}{\sqrt{2}} (1,r^2,{\bf 0}_T)$,
and $P_3 = \frac{M_B}{\sqrt{2}} (0,1-r^2,{\bf 0}_T)$, respectively,
where $r = M_{D_s}/M_B$ and we neglect the $K$ meson's mass $M_K$.
Putting the light (anti-)quark momenta in $B$, $D_s$ and
$K$ mesons as $k_1$, $k_2$, and $k_3$, respectively, we can
choose $k_1 = (x_1 P_1^+,0,{\bf k}_{1T})$,
$k_2 = (x_2 P_2^+,0,{\bf k}_{2T})$, and
$k_3 = (0, x_3 P_3^-,{\bf k}_{3T})$.
Then, integration over $k_1^-$, $k_2^-$, and $k_3^+$ in
eq.(\ref{eq:convolution1}) leads to
\begin{multline}
 \mbox{Amplitude}
\sim \int\!\!
d x_1 d x_2 d x_3
b_1 d b_1 b_2 d b_2 b_3 d b_3 \\
\mathrm{Tr} \bigl[ C(t) \Phi_B(x_1,b_1) \Phi_{D_s}(x_2,b_2)
\Phi_K(x_3, b_3) H(x_i, b_i, t) S_t(x_i)\, e^{-S(t)} \bigr],
\label{eq:convolution2}
\end{multline}
where $b_i$ is the conjugate space coordinate of $k_{iT}$, and
$t$ is the largest energy scale in $H$, as the function in terms of
$x_i$ and $b_i$.
The large logarithms ($\ln m_W/t$) coming from QCD radiative
corrections to four quark operators are included in the Wilson coefficients
$C(t)$.
The large double logarithms ($\ln^2 x_i$) on the longitudinal direction
are summed by the threshold resummation\cite{L3},
and they lead to $S_t(x_i)$ which smears the end-point singularities
on $x_i$.
The last term, $e^{-S(t)}$, contains two kinds of logarithms.
One of the large logarithms is due to the renormalization of
ultra-violet divergence $\ln tb$, the other is double logarithm
$\ln^2 b$ from the overlap of collinear and soft gluon corrections.
This Sudakov form factor suppresses the soft dynamics effectively \cite{soft}.
Thus it makes perturbative calculation of the hard part $H$ applicable
at intermediate scale, i.e., $M_B$ scale.
We calculate the $H$ for $B  \to D_s K$ decays in the first order in
$\alpha_s$ expansion and give the convoluted amplitudes in next section.

\section{Analytic formula}\label{sc:formula}

\subsection{The wave functions}
In order to calculate analytic formulas of the decay amplitude,
we use the wave functions $\Phi_{M,\alpha\beta}$ decomposed in terms of
spin structure.
In general, $\Phi_{M,\alpha\beta}$ having Dirac indices $\alpha,\beta$
are decomposed into 16 independent components, $1_{\alpha\beta}$,
$\gamma^\mu_{\alpha\beta}$, $\sigma^{\mu\nu}_{\alpha\beta}$,
$(\gamma^\mu\gamma_5)_{\alpha\beta}$, $\gamma_{5\alpha\beta}$.
If the considered meson $M$ is $B$ or $D_s$ meson,
to be pseudo-scalar and heavy meson, the structure
$(\gamma^\mu\gamma_5)_{\alpha\beta}$ and $\gamma_{5\alpha\beta}$
components remain as leading contributions.
Then, $\Phi_{M,\alpha\beta}$ is written by
\begin{equation}
 \Phi_{M,\alpha\beta} = \frac{i}{\sqrt{2N_c}}
\left\{
(\not \! P_M \gamma_5)_{\alpha\beta} \phi_M^A
+ \gamma_{5\alpha\beta} \phi_M^P
\right\},
\end{equation}
where $N_c = 3$ is color's degree of freedom,
$P_M$ is the corresponding meson's momentum, and
$\phi_M^{A,P}$ are Lorentz scalar wave functions.
As heavy quark effective theory leads to
$\phi_B^P \simeq M_B \phi_B^A$, then $B$ meson's wave function can be
expressed by
\begin{equation}
 \Phi_{B,\alpha\beta}(x,b) = \frac{i}{\sqrt{2N_c}}
\left[
(\not \! P_1 \gamma_5)_{\alpha\beta}
+ M_B \gamma_{5\alpha\beta}
\right]  \phi_B(x,b).
\end{equation}
According to ref.~\cite{Ball:1998je}, a pseudo-scalar meson moving fast
is parameterized by Lorentz scalar wave functions, $\phi$, $\phi_p$,
and $\phi_\sigma$ as
\begin{gather}
\langle D_s^-(P)|{\bar s}(z)\gamma_\mu \gamma_5 c(0)| 0 \rangle \simeq
- i f_{D_s} P_\mu\int_0^1\!\!\! dx\ e^{ix P z}\phi (x),
\label{pv} \\
\langle D_s^-(P)|{\bar s}(z)\gamma_5 c(0)|0 \rangle =
-if_{D_s} m_{0D_s} \int_0^1\!\!\! dx\ e^{ix P z}\phi_p(x),
\label{ps} \\
\langle D_s^-(P)|{\bar s}(z)\gamma_5 \sigma_{\mu\nu} c(0)|0 \rangle =
\frac{i}{6}f_{D_s} m_{0D_s} \left(1-\frac{M_{D_s}^2}{m_{0D_s}^2} \right)
(P_\mu z_\nu-P_\nu z_\mu)
\int_0^1\!\!\! dx\ e^{ix P z}\phi_\sigma(x),
\label{pt}
\end{gather}
where $m_{0D_s} = M_{D_s}^2/(m_c+m_s)$. We ignore the difference between
$c$ quark's mass and $D_s$ meson's mass in the perturbative
calculation.
This means, putting $\bar{\Lambda}' \equiv M_{D_s} - m_c$, the terms
proportional to $\bar{\Lambda}'/M_{D_s}$ are neglected.
In this approximation,
the contributions of eq.(\ref{pt}) are of higher power
than those of eqs.(\ref{pv}, \ref{ps}) by
${\cal O}\left( \frac{\bar{\Lambda}'}{M_{D_s}}\right)$
because of the factor $1- \frac{M_{D_s}^2}{m_{0D_s}^2}$ in
eq.(\ref{pt}), and we neglect the $\gamma_5 \sigma_{\mu\nu}$ component
in the $D_s$ meson's wave function.
In addition, the eq.(\ref{pv}), eq.(\ref{ps}), and the relations
\begin{gather}
\frac{\partial\ }{\partial z_\mu}
\langle D_s^-(P)|{\bar s}(z)\gamma_\mu \gamma_5 c(0)
| 0 \rangle =
i\, m_s \langle D_s^-(P)|{\bar s}(z)\gamma_5 c(0)|0 \rangle,  \\
\frac{\partial\ }{\partial z_\mu}
\langle D_s^-(P)|{\bar s}(0)\gamma_\mu \gamma_5 c(z)
| 0 \rangle =
i\, m_c \langle D_s^-(P)|{\bar s}(0)\gamma_5 c(z)|0 \rangle,
\end{gather}
with equations of motion lead to
\begin{equation}
\phi_p(x) = \phi(x) +
{\cal O}\left( \frac{\bar{\Lambda}'}{M_{D_s}}\right).
\end{equation}
Therefore
the $D_s$ meson's wave function can be expressed by one
Lorentz scalar wave function,
\begin{equation}
 \Phi_{D_s,\alpha\beta}(x,b) = \frac{i}{\sqrt{2N_c}}
\left[
(\gamma_5 \not \! P_2 )_{\alpha\beta}
+ M_{D_s} \gamma_{5\alpha\beta}
\right]  \phi_{D_s}(x,b),
\end{equation}
where $\phi_{D_s}$ is defined by
\begin{equation}
 \phi_{D_s}(x) = \frac{f_{D_s}}{2\sqrt{2N_c}}\phi(x)
 = \frac{f_{D_s}}{2\sqrt{2N_c}}\phi_p(x).
\end{equation}
The wave function $\phi_M$ for $M = B, D_s$ meson is normalized by its
decay constant $f_M$
\begin{equation}
 \int_0^1 \!\! dx\  \phi_M (x, b=0)
= \frac{f_M}{2\sqrt{2N_c}}.
\label{eq:normalization}
\end{equation}

In contrast to the $B$ and $D$ mesons,
for the $K$ meson, being light meson,
the $\gamma_5 \sigma^{\mu\nu}$ component remains
because the factor corresponding to $1- \frac{M_{D_s}^2}{m_{0D_s}^2}$ in
eq.(\ref{pt}) is ${\cal O}(1)$.
Then, $K$ meson's wave function is parameterized by
Lorentz scalar wave functions $\phi_K^{A,P,T}$ as
\begin{align}
 \Phi_{K,\alpha\beta}(x_3,b_3) &=
P_3^- \int \! \frac{dz^+}{2\pi} e^{-i k_3 \cdot z}
\langle K^+(P_3)| {\bar u}_{\beta i}(z) s_{\alpha j}(0)|0\rangle
\nonumber \\
&= \frac{i\delta_{ij}}{\sqrt{2N_c}}
\Bigl[ \gamma_5 \not \! P_3 \phi_K^A(x_3,b_3)
+ m_{0K} \gamma_5 \phi_K^P(x_3,b_3) \nonumber \\
& \qquad\qquad\qquad\qquad\qquad
 + m_{0K} \gamma_5 (\not v \not n - 1)\phi_K^T(x_3,b_3)
\Bigr]_{\alpha\beta}
\end{align}
where $v = (0,1,{\bf 0}_T ) \propto P_3$,
$n = (1,0,{\bf 0}_T) \propto z$, and
$m_{0K} = M_K^2/(m_u + m_s)$.
In the numerical analysis we will use $\phi_K^{A,P,T}$ which were
calculated from QCD sum rule \cite{Ball:1998je}.
They will be shown in section \ref{sc:neval}.

\subsection{$B^0 \to D_s^- K^+$ decay}

We first consider the neutral $B^0$ decay $B^0 \to D_s^- K^+$.
The effective Hamiltonian at the scale lower than $M_W$ related to
this decay is given as \cite{Buchalla:1996vs}
\begin{gather}
 H_\mathrm{eff} = \frac{G_F}{\sqrt{2}} V_{cb}^*V_{ud} \left[
C_1(\mu) O_1(\mu) + C_2(\mu) O_2(\mu) \right], \\
  O_1 = (\bar{b}d)_{V-A} (\bar{u}c)_{V-A}, \quad
 O_2 = (\bar{b}c)_{V-A} (\bar{u}d)_{V-A},
\end{gather}
where $C_{1,2}(\mu)$ are Wilson coefficients at renormalization scale
$\mu$, and summation in $\mathrm{SU}(3)_c$ color's index $\alpha$ and
chiral projection,
$\sum_\alpha \bar{q}_\alpha \gamma^\nu(1-\gamma_5)q'_\alpha$,
are abbreviated to $(\bar{q}q')_{V-A}$.
The lowest order diagrams contributing to $B^0 \to D_s^- K^+$ are
drawn in Fig.\ref{fig:diagrams1} according to this effective
Hamiltonian.
As stated above, $B \to D_s K$ decays only have  annihilation
diagrams.

We get the following analytic formulas by calculating the hard part $H$
at first order in $\alpha_s$.
Together with the meson wave functions,
the amplitude for the factorizable annihilation diagram
in Fig.\ref{fig:diagrams1}(a) and (b)
results in $F_a^{(i=2)}$,
\begin{multline}
F_a^{(i)} = -16\pi C_F M_B^2 \int_0^1\!\!\! dx_2 dx_3
 \int_0^\infty\!\!\!\!\!  b_2 db_2\, b_3 db_3\ \phi_{D_s}(x_2,b_2) \\
\times \Bigl[ \bigl\{
(1-r^2)\left( 1 - 2 r^2 - (1-r^2)x_3 \right) \phi_K^A(x_3,b_3) \\
+ r \left( 3 - r^2 - 2 (1-r^2)x_3 \right) r_K \phi_K^P(x_3,b_3) \\
- r (1-r^2) (1 - 2 x_3 ) r_K \phi_K^T(x_3,b_3)
\bigr\} E_{f}^i(t_a^1) h_a(x_2,x_3,b_2,b_3) \\
- \bigl\{
(1-r^2)x_2\phi_K^A(x_3,b_3) \\
+ 2 r (1-r^2+x_2) r_K \phi_K^P(x_3,b_3)
\bigr\}
E_{f}^i(t_a^2) h_a(1-x_3,1-x_2,b_3,b_2) \Bigr],
\label{eq:Fa}
\end{multline}
where $C_F = 4/3$ is the group factor of $\mathrm{SU}(3)_c$ gauge group,
and $r_K = m_{0K}/M_B$,
and the functions $E_{f}^i$, $t_a^{1,2}$, $h_a$ are given in the appendix.
The explicit form for the wave functions, $\phi_M$,
is given in the next section.
The amplitude for the nonfactorizable annihilation diagram in
Fig.\ref{fig:diagrams1}(c) and (d) results in
\begin{multline}
M_a  =  \frac{1}{\sqrt{2N_c}} 64\pi C_F M_B^2
\int_0^1\!\!\! dx_1 dx_2 dx_3
 \int_0^\infty\!\!\!\!\! b_1 db_1\, b_2 db_2\
\phi_B(x_1,b_1) \phi_{D_s}(x_2,b_2) \\
\times \Bigl[
\bigl\{ (1-r^2) \left((1-r^2)(1 - x_3) + r^2 x_2 \right) \phi_K^A(x_3,b_2) \\
 + r \left(x_2 + (1-r^2)(1-x_3) \right) r_K \phi_K^P(x_3,b_2) \\
 + r \left(x_2 - (1-r^2)(1-x_3) \right) r_K \phi_K^T(x_3,b_2)
\bigr\}
E_{m}(t_{m}^1) h_a^{(1)}(x_1, x_2,x_3,b_1,b_2) \\
- \bigl\{
(1-r^2) \left( (1+r^2)x_2 -r^2 \right) \phi_K^A(x_3,b_2) \\
 + r (2 + x_2 + (1-r^2)(1-x_3)) r_K \phi_K^P(x_3,b_2) \\
 + r (-x_2 + (1-r^2)(1-x_3)) r_K \phi_K^T(x_3,b_2)
\bigr\}
E_{m}(t_{m}^2) h_a^{(2)}(x_1, x_2,x_3,b_1,b_2)
\Bigr],
\label{eq:Ma1}
\end{multline}
where $x_1$ dependence in the numerators of the hard part are neglected
by the assumption $x_1 \ll x_2, x_3$\footnote{
We don't apply this approximation to the denominators of the propagator
which are sensitive to the variable $x_1$.
Because such a $x_1$ behaves as the cut off,
the resultant branching ratio is smaller than it given in
ref.~\cite{Ukai:2001yp} where
the whole $x_1$ in the hard part are neglected.}.
The total decay amplitude for $B^0 \to D_s^- K^+$ decay is given
as
\begin{equation}
 A = f_B F_a^{(2)}  + M_a , \label{eq:neut_amp}
\end{equation}
where the overall factor is included in the decay width
with the kinematics factor.
The decay width is expressed as
\begin{equation}
 \Gamma(B^0 \to D_s^- K^+) = \frac{G_F^2 M_B^3}{128\pi} (1-r^2)
|V_{cb}^*V_{ud} A|^2.
\label{eq:neut_width}
\end{equation}
The decay width for CP conjugated mode, $\overline{B}^0 \to D_s^+ K^-$,
is the same value as $B^0 \to D_s^- K^+$, just replacing
$V_{cb}^*V_{ud}$ with $V_{cb}V_{ud}^*$.
Since there is only one kind of CKM phase involved in the decay, there is
no CP violation in the standard model.

\subsection{$B^+ \to D_s^+ \overline{K}^0$ decay}

The effective Hamiltonian related to
$B^+ \to D_s^+ \overline{K}^0$ decay is given as
\begin{gather}
 H_\mathrm{eff} = \frac{G_F}{\sqrt{2}} V_{ub}^*V_{cd} \left[
C_1(\mu) O_1(\mu) + C_2(\mu) O_2(\mu) \right], \\
  O_1 = (\bar{b}d)_{V-A} (\bar{c}u)_{V-A}, \quad
 O_2 = (\bar{b}u)_{V-A} (\bar{c}d)_{V-A}.
\end{gather}
The amplitude for the factorizable annihilation diagram results in
$-F_a^{(i=1)}$.
The amplitude for the nonfactorizable annihilation diagram results in
\begin{multline}
M_a^{\prime} =  \frac{1}{\sqrt{2N_c}} 64\pi C_F M_B^2
\int_0^1\!\!\! dx_1 dx_2 dx_3
 \int_0^\infty\!\!\!\!\! b_1 db_1\, b_2 db_2\
 \phi_B(x_1,b_1) \phi_{D_s}(x_2,b_2) \\
\times \Bigl[ \bigl\{ (1-r^4)x_2 \phi_K^A(x_3,b_2)
 + r \left(x_2+(1-r^2)(1-x_3)\right) r_K \phi_K^P(x_3,b_2) \\
 + r \left(-x_2+(1-r^2)(1-x_3)\right) r_K \phi_K^T(x_3,b_2)
\bigr\}
E_{m}'(t_{m}^1) h_a^{(1)}(x_1, x_2,x_3,b_1,b_2) \\
- \bigl\{
(1-r^2) \left( (1-r^2)(1-x_3) - r^2+r^2 x_2 \right) \phi_K^A(x_3,b_2) \\
 + r \left(2 + x_2 + (1-r^2)(1-x_3) \right) r_K \phi_K^P(x_3,b_2) \\
 + r \left(x_2 - (1-r^2)(1-x_3) \right) r_K \phi_K^T(x_3,b_2)
\bigr\}E_{m}'(t_{m}^2) h_a^{(2)}(x_1, x_2,x_3,b_1,b_2)
\Bigr].
\label{eq:Ma2}
\end{multline}
Thus, the total decay amplitude $A'$ and decay width $\Gamma$ for
$B^+ \to D_s^+ \overline{K}^0$ decay is given as
\begin{gather}
  A' = - f_B F_a^{(1)} + M_a^{\prime},
\label{eq:chrg_amp} \\
 \Gamma(B^+ \to D_s^+ \overline{K}^0) = \frac{G_F^2 M_B^3}{128\pi} (1-r^2)
|V_{ub}^*V_{cd} A'|^2 .
\label{eq:chrg_width}
\end{gather}
The decay width for CP conjugated mode, $B^- \to D_s^- K^0$,
is the same value as $B^+ \to D_s^+ \overline{K}^0$.
Similar to the $B^0$ decay, there is also no CP violation in this decay within
standard model.

\section{Numerical evaluation}\label{sc:neval}

In this section we show numerical results obtained from the previous
formulas.
At the beginning, we give the branching ratios predicted from the same
parameters and wave functions that are adopted in the other works.
Secondly, we show the theoretical errors due to uncertainty of some
parameters.

For the $B$  meson's wave function, there is a sharp peak at the small $x$
region, we use
\begin{equation}
\phi_B(x,b) = N_B x^2(1-x)^2 \exp \left[
-\frac{M_B^2\ x^2}{2 \omega_b^2} -\frac{1}{2} (\omega_b b)^2
\right],
\end{equation}
which is adopted in ref. \cite{Keum:2000,Lu:2000em}.
This choice of $B$ meson's wave function is almost a best  fit from
the
$B\to K\pi$, $\pi \pi$ decays.
For the $D_s$ meson's wave function,
we assume the same form as $D$ meson's
one except for the normalization \cite{h2h}
\begin{equation}
\phi_{D_s}(x) = \frac{3}{\sqrt{2 N_c}} f_{D_s}
x(1-x)\{ 1 + a_{D_s} (1 -2x) \}.
\end{equation}
Since $c$ quark is much heavier than $s$ quark, this function is
peaked at $c$ quark side, i.e. small $x$ region, too.
The wave functions $\phi_K^{A,P,T}$ of the $K$ meson are expanded by
Gegenbauer polynomials,
\begin{eqnarray}
\phi_K^A(x) &=& \frac{f_K}{2\sqrt{2 N_c}} 6 x(1-x)
\left\{ 1 - a_1^K \cdot 3 \xi +
a_2^K \cdot \frac{3}{2} (-1 + 5 \xi^2) \right\}, \\
\phi_K^P(x) &=& \frac{f_K}{2\sqrt{2 N_c}}
\left\{ 1 + a_{p1}^K \cdot \frac{1}{2} (-1 + 3\xi^2)
+ a_{p2}^K \cdot \frac{1}{8} (3 - 30\xi^2 +35\xi^4) \right\}, \\
\phi_K^T(x) &=& \frac{f_K}{2\sqrt{2 N_c}} (1-2x)
\left\{ 1 + a_T^K \cdot 3(-3 + 5 \xi^2) \right\},
\end{eqnarray}
where $\xi = 2 x -1$.
The coefficients $a^K_{1, 2, p1, p2, T}$ calculated from QCD sum rule
are given in ref.\cite{Ball:1998je}, and their values are
\begin{equation}
a_1^K = 0.17,\quad  a_2^K = 0.2,\quad  a_{p1}^K = 0.212,
 \quad a_{p2}^K = -0.148,\quad a_T^K = 0.0527,
\label{eq:parm_phiK}
\end{equation}
for $m_{0K} = 1.6 \mbox{ GeV}$.
In addition, we use the following input parameters:
\begin{gather}
 M_B = 5.28 \mbox{ GeV},\  M_{D_s} = 1.969 \mbox{ GeV},
\label{eq:parm1} \\
f_B = 190 \mbox{ MeV},\  f_K = 160 \mbox{ MeV},\
f_{D_s} = 241 \mbox{ MeV}, \\
m_{0K} = 1.6 \mbox{ GeV},\ \omega_b = 0.4 \mbox{ GeV},\  a_{D_s} = 0.3.
\label{eq:shapewv}
\end{gather}
With these values  and eq.(\ref{eq:normalization})
we get the normalization factor $N_B = 91.745$ GeV.
We show the decay amplitudes calculated with the above parameters at Table
\ref{tb:amplitudes}.
For the neutral decay $B^0 \to D_s^- K^+$, the dominant contribution is
the nonfactorizable annihilation diagrams, where the contribution $M_a$
 is proportional
to the Wilson coefficient $C_2(t)$, which is of order one. The factorizable
annihilation diagram contribution is proportional to $a_2=C_1+C_2/3$,
which is one order magnitude smaller.
For the charged decay $B^+ \to D_s^+\overline{K}^0$, it is the inverse
situation. The Wilson coefficient in $M_a'$ is $C_1(t)$, which is smaller
than the one in $F_a^{(1)}$, $a_1=C_1/3 +C_2$.
 Both amplitudes $A$ and $A'$ are at the same order magnitude here.

The propagators of inner quark and gluon in Figure 2 and 3 are usually
proportional to  $1/x_i$.
One may suspect that these amplitudes are enhanced by the endpoint
singularity around $x_i \sim 0$. This can be explicitly found in
eq.(\ref{eq:propagator1}, \ref{eq:propagator2}),
where the Bessel function $\mathrm{Y}_0$ diverges at
$x_i \sim 0$ or $1$.
However this is not the case in
our calculation. First we introduce the transverse momentum of
quark, such that the propagators become $1/(x_i x_j +k_T^2)$.
Secondly, the Sudakov form factor $\mathrm{Exp}[-S]$ suppresses the region
of small $k^2_T$. Therefore there is no singularity in our
calculation. The dominant contribution is not from the endpoint of the
wave function. As a proof,
in our numerical calculations, for example, an expectation value of $\alpha_s$
in the integration for $M_a$ results in
$\langle \alpha_s/\pi \rangle = 0.10$, which gives the dominant
contribution to $B^0 \to D_s^- K^+$ decay.
Therefore, the perturbative calculations are self-consistent.
\begin{table}[htbp]
 \begin{center}
  \begin{tabular}[b]{cr|cr}
   \hline
   \hline
   \multicolumn{2}{c|}{$B^0 \to D_s^-K^+$} &
   \multicolumn{2}{c}{$B^+ \to D_s^+ \overline{K}^0$} \\
   \hline \hline
   $f_B F_a^{(2)}$ & $-0.84 +1.57\, i$ &
   $- f_B F_a^{(1)}$ & $-17.16 + 6.59\, i$ \\
   $M_a$ & $1.99  -18.36\, i$ &
   $M_a^{\prime}$ & $9.37 + 8.34\, i$ \\
   \hline
   $A$ & $1.15 -16.79\, i$ &
   $A'$ & $-7.80 +14.93\, i$        \\
    \hline
   \hline
  \end{tabular}
 \end{center}
 \caption{Amplitudes($10^{-3}$ GeV) with parameters eqs.
 (\ref{eq:parm1}-\ref{eq:shapewv}).}
 \label{tb:amplitudes}
\end{table}

Now we can calculate the branching ratio according to eqs.
(\ref{eq:neut_amp}, \ref{eq:neut_width}, \ref{eq:chrg_amp},
\ref{eq:chrg_width}).
Here we use CKM matrix elements and the life times\cite{pdg},
\begin{gather}
|V_{ud}|=0.9734\pm 0.0008, \  |V_{ub}|= (3.6 \pm 0.7)\times 10^{-3}, \\
|V_{cb}|= (41.2\pm 2.0)\times 10^{-3}, \  |V_{cd}|=0.224\pm 0.016,
\label{eq:KMmatrix} \\
 \tau_{B^\pm}=1.67\times 10^{-12}\mbox{ s,}\quad
 \tau_{B^0}=1.54\times 10^{-12}\mbox{ s}.
\end{gather}
The predicted branching ratios are
\begin{gather}
 \mathrm{Br}(B^0 \to D_s^- K^+) = 4.57 \times 10^{-5},
\label{eq:br1}\\
 \mathrm{Br}(B^+ \to D_s^+ \overline{K}^0) = 2.01 \times 10^{-8}.
\label{eq:br2}
\end{gather}
The $B^0 \to D_s^- K^+$ decay is observed at Belle\cite{Krokovny:2002pe}
and BaBar\cite{Aubert:2002eu},
\begin{align}
 \mathrm{Br}(B^0 \to D_s^- K^+)
& = (4.6{}^{+1.2}_{-1.1}\pm 1.3) \times 10^{-5}, & & \mbox{Belle, } \\
 \mathrm{Br}(B^0 \to D_s^- K^+)
& = (3.2\pm 1.0 \pm 1.0) \times 10^{-5}, & & \mbox{BaBar. }
\end{align}
For the $B^+ \to D_s^+ \overline{K}^0$ decay, there is only  upper limit given
at $90$\% confidence level\cite{pdg},
\begin{equation}
 \mathrm{Br}(B^+ \to D_s^+ \overline{K}^0) < 1.1 \times 10^{-3}.
\end{equation}
It is easy to see that our results are consistent with the data.

Despite the calculated perturbative annihilation contributions,
there is also hadronic picture for the $B^0 \to D_s^- K^+$ decay:
$B^0 \to D^- \pi^+(\rho^+)$ then $D^- \pi^+ (\rho^+) \to D_s^- K^+$ through final
state interaction.  Our numerical results show that the PQCD
contribution to this decay is already enough to account for the
experimental measurement. It implies that the soft final state
interaction is      not important in the  $B^0 \to D_s^- K^+$
decay. This is consistent with the argument in ref.\cite{cl}.

The branching ratios obtained from the analytic formulas may be sensitive
to various parameters,
such as  parameters in eqs.(\ref{eq:parm_phiK}),
eqs.(\ref{eq:shapewv}).
It is important to give the limits of the branching ratio
when we choose the parameters to some extent.
Table~\ref{tb:sensitivity} shows the sensitivity of the branching ratio to
$30$\% change of parameters in eqs.(\ref{eq:parm_phiK}),
eqs.(\ref{eq:shapewv}).
\begin{table}[htbp]
\begin{center}
\begin{tabular}[t]{r|cc}
 \hline     \hline
& $(10^{-5})$ & $(10^{-8})$ \\
 $a_1^K$ & $\mathrm{Br}(B^0 \to D_s^- K^+)$ &
$\mathrm{Br}(B^+ \to D_s^+ \overline{K}^0)$ \\
 \hline
 $0.119$ & $5.01$ & $2.32$ \\
 $0.170$ & $4.57$ & $2.01$ \\
 $0.221$ & $4.28$ & $1.73$ \\
 \hline
 \hline
 $a_2^K$ & $\mathrm{Br}(B^0 \to D_s^- K^+)$ &
$\mathrm{Br}(B^+ \to D_s^+ \overline{K}^0)$ \\
 \hline
 $0.14$ & $4.38$ & $1.74$ \\
 $0.20$ & $4.57$ & $2.01$ \\
 $0.26$ & $4.93$ & $2.42$ \\
 \hline
 \hline
 $m_{0K}$ & $\mathrm{Br}(B^0 \to D_s^- K^+)$ &
$\mathrm{Br}(B^+ \to D_s^+ \overline{K}^0)$ \\
 \hline
 $1.12$ & $4.36$ & $1.61$ \\
 $1.60$ & $4.57$ & $2.01$ \\
 $2.08$ & $5.18$ & $2.53$ \\
 \hline
 \hline
 $\omega_b$ & $\mathrm{Br}(B^0 \to D_s^- K^+)$ &
$\mathrm{Br}(B^+ \to D_s^+ \overline{K}^0)$ \\
 \hline
 $0.28$ & $6.36$ & $2.39$ \\
 $0.40$ & $4.57$ & $2.01$ \\
 $0.52$ & $3.34$ & $1.77$ \\
 \hline
 \hline
 $a_{D_s}$ & $\mathrm{Br}(B^0 \to D_s^- K^+)$ &
$\mathrm{Br}(B^+ \to D_s^+ \overline{K}^0)$ \\
 \hline
 $0.21$ & $4.38$ & $1.87$ \\
 $0.30$ & $4.57$ & $2.01$ \\
 $0.39$ & $4.95$ & $2.19$     \\
  \hline          \hline
\end{tabular}
\end{center}
\caption{The sensitivity of the branching ratio to
$30$\% change of parameters in eqs.(\ref{eq:parm_phiK}),
eqs.(\ref{eq:shapewv}).
Here we don't present the sensitivity to $a_{p1,p2,T}^K$ because
the branching ratios are insensitive to them.}
\label{tb:sensitivity}
\end{table}
It is found that uncertainty of the predictions on PQCD is mainly due to
 $m_{0K}$ and $\omega_b$.
Below we show the limits of the branching ratio within the suitable
 ranges on $m_{0K}$ and $\omega_b$.
The appropriate extent of $m_{0K}$ can be
found in ref.\cite{Ball:1998je},
\begin{equation}
 1.4 \mbox{ GeV} \leq m_{0K} \leq 1.8 \mbox{ GeV}.
\label{eq:m0K}
\end{equation}
$a^K_{p1, p2, t}$ in the wave functions $\phi_K^{P,T}$ are given as
functions with respect to $m_{0K}$, $a_2^K$, and some input parameters,
$\eta_3$, $\omega_3$ in ref.\cite{Ball:1998je}.
Within eq.~(\ref{eq:m0K}),
the branching ratios normalized by the decay constants and the CKM
matrix elements result in
\begin{gather}
\mathrm{Br}(B^0 \to D_s^- K^+)
= (4.57_{-0.10}^{+0.26}) \times 10^{-5}
\left( \frac{f_B\ f_{D_s}}{190\mbox{ MeV}\cdot 241\mbox{ MeV}} \right)^2
\left(
\frac{|V_{cb}^*\ V_{ud}|}
{0.0412\cdot 0.9734}
\right)^2
, \\
\mathrm{Br}(B^+ \to D_s^+ \overline{K}^0)
= (2.01_{-0.18}^{+0.16}) \times 10^{-8}
\left( \frac{f_B\ f_{D_s}}{190\mbox{ MeV}\cdot 241\mbox{ MeV}} \right)^2
\left(
\frac{|V_{ub}^*\ V_{cd}|}
{ 0.0036 \cdot 0.224}
\right)^2
.
\end{gather}
 From the $B\to K$ transition form factor $f_+^K(0)$,
the appropriate extent of $\omega_b$ can be obtained.
$f_+^K(0)$ calculated from PQCD at $m_{0K}= 1.6$ GeV in the region
\begin{equation}
 0.35 \mbox{ GeV} \leq \omega_b \leq 0.46 \mbox{ GeV},
\label{eq:wb}
\end{equation}
is consistent with $f_+^K(0)$ by QCD sum rules given in
ref.~\cite{Ball:1998je}.
The branching ratios calculated at the region of eqs.~(\ref{eq:wb})
are found within
\begin{gather}
\mathrm{Br}(B^0 \to D_s^- K^+)
= (4.57_{-0.59}^{+0.77}) \times 10^{-5}
\left( \frac{f_B\ f_{D_s}}{190\mbox{ MeV}\cdot 241\mbox{ MeV}} \right)^2
\left(
\frac{|V_{cb}^*\ V_{ud}|}
{0.0412\cdot 0.9734}
\right)^2
, \\
\mathrm{Br}(B^+ \to D_s^+ \overline{K}^0)
= (2.01_{-0.10}^{+0.20}) \times 10^{-8}
\left( \frac{f_B\ f_{D_s}}{190\mbox{ MeV}\cdot 241\mbox{ MeV}} \right)^2
\left(
\frac{|V_{ub}^*\ V_{cd}|}
{ 0.0036 \cdot 0.224}
\right)^2
.
\end{gather}
In the factorizable contribution, the $B$ meson's wave function is
integrated and normalized by the decay constant $f_B$.
Thus, in the factorizable dominant decay,
$B^+ \to D_s^+ \overline{K}^0$,
its branching ratio is insensitive to the change of $\omega_b$.


\section{Conclusion}\label{sc:concl}

In two-body hadronic $B$ meson decays, the final state mesons are
moving very fast, since each of them carry more than 2 GeV energy.
There is not enough time for them to exchange soft gluons.
The soft final state interaction is not important in the two-body
$B$ decays. This is consistent with the argument based on
color-transparency \cite{bjo}.
We thus neglect the soft final state interaction in the PQCD approach.
The PQCD with Sudakov form factor is a self-consistent approach to
describe the two-body $B$ meson decays.
Although the  annihilation diagrams are suppressed comparing to
other spectator diagrams, but their contributions are not
negligible in PQCD approach \cite{Keum:2000,Lu:2000em}.

In this paper, we calculate the $B^0 \to D_s^- K^+$ and
$B^+ \to D_s^+ \overline{K}^0$ decays, which  occur purely via
annihilation type diagrams.
The branching ratios are still sizable.
The $B^0 \to D_s^- K^+$ decay has been observed in the $B$
factories~\cite{Krokovny:2002pe,Aubert:2002eu}.
This is the first channel  measured
in $B$ decays via annihilation type diagram.
The fact that the predicted branching ratio is in good agreement
with the data means that PQCD approach gives the annihilation amplitude
correctly,
the soft final state interaction is probably small in $B$ decays,
and it is one of the evidences to justify PQCD approach.

\section*{Acknowledgments}

We thank our PQCD group members:
C.-H. Chen, Y.-Y. Keum, E. Kou, T. Kurimoto, H.-n. Li, T. Morozumi, 
S. Mishima, M. Nagashima, A.I. Sanda, N. Sinha, R. Sinha, M.Z. Yang and
T. Yoshikawa for fruitful discussions.
This work is partly supported by National Science Foundation of
China under Grant (No. 90103013 and 10135060).
The research of K.U. is supported by the Japan Society for Promotion of
Science under the Predoctoral Research Program. 
This work was supported by a Grand-in Aid for Special Project Research 
(Physics of CP violation) and by a Grand-in-Aid for Science Research from
the Ministry of Education, Culture, Sports, Science and Technology, 
Japan (No. 12004276).

\begin{appendix}

\section{Some functions}
The definitions of some functions used in the text are presented
in this appendix.
In the numerical analysis we use one loop expression for strong coupling
constant,
\begin{equation}
 \alpha_s (\mu) = \frac{4 \pi}{\beta_0 \log (\mu^2 / \Lambda^2)},
\label{eq:alphas}
\end{equation}
where $\beta_0 = (33-2n_f)/3$ and $n_f$ is number of active flavor at
appropriate scale. $\Lambda$ is QCD scale, which we use as $250$ MeV at
$n_f=4$.
We also use leading logarithms expressions for Wilson coefficients
$C_{1,2}$ presented in ref.\cite{Buchalla:1996vs}.
Then, we put $m_t = 170$ GeV, $m_W = 80.2$ GeV, and $m_b = 4.8$ GeV.

The function $E_f^i$, $E_m$, and $E'_m$
including Wilson coefficients
are defined as
\begin{gather}
 E_{f}^i(t) = a_i(t) \alpha_s(t)\, e^{-S_D(t)-S_K(t)}, \\
 E_{m}(t) = C_2(t) \alpha_s(t)\, e^{-S_B(t)-S_D(t)-S_K(t)}, \\
 E_{m}'(t) = C_1(t) \alpha_s(t)\, e^{-S_B(t)-S_D(t)-S_K(t)},
\end{gather}
 where
\begin{equation}
 a_1(t) = \frac{C_1(t)}{N_c} + C_2(t),\quad
 a_2(t) = C_1(t) + \frac{C_2(t)}{N_c},
\end{equation}
and $S_B$, $S_D$, and $S_K$ result from summing both double logarithms
caused by soft gluon corrections and single ones due to
the renormalization of ultra-violet divergence.
The above $S_{B, D, K}$ are defined as
\begin{gather}
S_B(t) = s(x_1P_1^+,b_1) +
2 \int_{1/b_1}^t \frac{d\mu'}{\mu'} \gamma_q(\mu'), \\
S_D(t) = s(x_2P_2^+,b_3) +
2 \int_{1/b_2}^t \frac{d\mu'}{\mu'} \gamma_q(\mu'), \\
S_K(t) = s(x_3P_3^+,b_3) + s((1-x_3)P_3^+,b_3) +
2 \int_{1/b_3}^t \frac{d\mu'}{\mu'} \gamma_q(\mu'),
\end{gather}
where $s(Q,b)$, so-called Sudakov factor, is given as
\cite{Li:1999kn}
\begin{eqnarray}
  s(Q,b) &=& \int_{1/b}^Q \!\! \frac{d\mu'}{\mu'} \left[
 \left\{ \frac{2}{3}(2 \gamma_E - 1 - \log 2) + C_F \log \frac{Q}{\mu'}
 \right\} \frac{\alpha_s(\mu')}{\pi} \right. \nonumber \\
& &  \left.+ \left\{ \frac{67}{9} - \frac{\pi^2}{3} - \frac{10}{27} n_f
 + \frac{2}{3} \beta_0 \log \frac{\gamma_E}{2} \right\}
 \left( \frac{\alpha_s(\mu')}{\pi} \right)^2 \log \frac{Q}{\mu'}
 \right],
 \label{eq:SudakovExpress}
\end{eqnarray}
 $\gamma_E=0.57722\cdots$ is Euler constant,
and $\gamma_q = \alpha_s/\pi$ is the quark anomalous dimension.

The functions $h_a$, $h_a^{(1)}$, and $h_a^{(2)}$  in the decay
amplitudes consist of two parts: one is the jet function $S_t(x_i)$
derived by the threshold resummation\cite{L3},
the other is the propagator of virtual quark and gluon.
They are defined by
\begin{align}
& h_a(x_2,x_3,b_2,b_3) = S_t(1-x_3)\left( \frac{\pi i}{2}\right)^2
H_0^{(1)}(M_B\sqrt{(1-r^2)x_2(1-x_3)}\, b_2) \nonumber \\
&\times \left\{
H_0^{(1)}(M_B\sqrt{(1-r^2)(1-x_3)}\, b_2)
J_0(M_B\sqrt{(1-r^2)(1-x_3)}\, b_3)
\theta(b_2 - b_3) + (b_2 \leftrightarrow b_3 ) \right\},
\label{eq:propagator1} \\
& h^{(j)}_a(x_1,x_2,x_3,b_1,b_2) = \nonumber \\
& \biggl\{
\frac{\pi i}{2} \mathrm{H}_0^{(1)}(M_B\sqrt{(1-r^2)x_2(1-x_3)}\, b_1)
 \mathrm{J}_0(M_B\sqrt{(1-r^2)x_2(1-x_3)}\, b_2) \theta(b_1-b_2)
\nonumber \\
& \qquad\qquad\qquad\qquad + (b_1 \leftrightarrow b_2) \biggr\}
 \times\left(
\begin{matrix}
 \mathrm{K}_0(M_B F_{(j)} b_1), & \text{for}\quad F^2_{(j)}>0 \\
 \frac{\pi i}{2} \mathrm{H}_0^{(1)}(M_B\sqrt{|F^2_{(j)}|}\ b_1), &
 \text{for}\quad F^2_{(j)}<0
\end{matrix}\right),
\label{eq:propagator2}
\end{align}
where $\mathrm{H}_0^{(1)}(z) = \mathrm{J}_0(z) + i\, \mathrm{Y}_0(z)$, and
$F_{(j)}$s are defined by
\begin{equation}
 F^2_{(1)} = (1-r^2)(x_1 -x_2)(1- x_3),\
F^2_{(2)} = x_1 +x_2+(1-r^2)(1-x_1-x_2)(1-x_3).
\end{equation}
We adopt the parametrization for $S_t(x)$ of the factorizable
contributions,
\begin{equation}
 S_t(x) = \frac{2^{1+2c}\Gamma(3/2 +c)}{\sqrt{\pi} \Gamma(1+c)}
[x(1-x)]^c,\quad c = 0.3,
\end{equation}
which is proposed in ref.~\cite{Kurimoto:2001zj}.
In the nonfactorizable annihilation contributions,
$S_t(x)$ gives a very small numerical effect to the amplitude\cite{L4}.
Therefore, we drop $S_t(x)$ in $h_a^{(1)}$ and $h_a^{(2)}$.
The hard scale $t$'s in the amplitudes are taken as the largest energy
scale in the $H$ to kill the large logarithmic radiative corrections:
\begin{gather}
 t_a^1 = \mathrm{max}(M_B \sqrt{(1-r^2)(1-x_3)},1/b_2,1/b_3), \\
 t_a^2 = \mathrm{max}(M_B \sqrt{(1-r^2)x_2},1/b_2,1/b_3), \\
 t_{m}^j = \mathrm{max}(M_B \sqrt{|F^2_{(j)}|},
M_B \sqrt{(1-r^2)x_2(1-x_3) }, 1/b_1,1/b_2).
\end{gather}

\end{appendix}

\begin{figure}[htbp]
\begin{flushleft}
{\bf\Large Figure Captions}
\end{flushleft}
 \caption{$D_s K$ are produced only by annihilation between $\bar{b}$ and $d$
quark in the $B $ meson from point of view of quark model.
While $W$ boson exchange causes $\bar{b}d \to \bar{c}u $,
the $\bar{s}s$ quarks included in $D_s K$ are produced from a hard gluon.}
 \label{fig:need1gluon}
 \caption{Diagrams for $B^0 \to D_s^- K^+$ decay. The factorizable
 diagrams (a),(b) contribute to $F_a^{(2)}$, and nonfactorizable (c),
 (d) do to $M_a$.}
 \label{fig:diagrams1}
 \caption{Diagrams for $B^+ \to D_s^+ \overline{K}^0$ decay. The factorizable
 diagrams (a),(b) contribute to $F_a^{(1)}$, and nonfactorizable (c),
 (d) do to $M_a^{\prime}$.}
 \label{fig:diagrams2}
\end{figure}

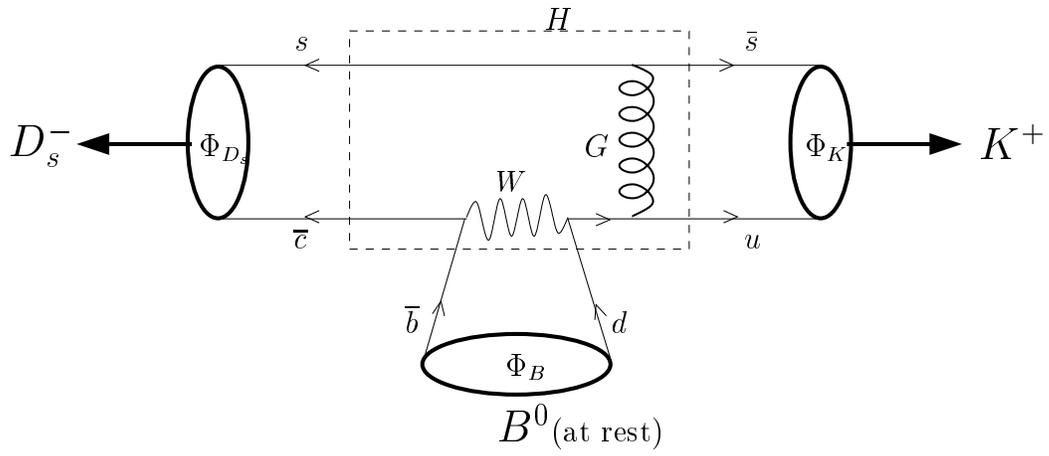
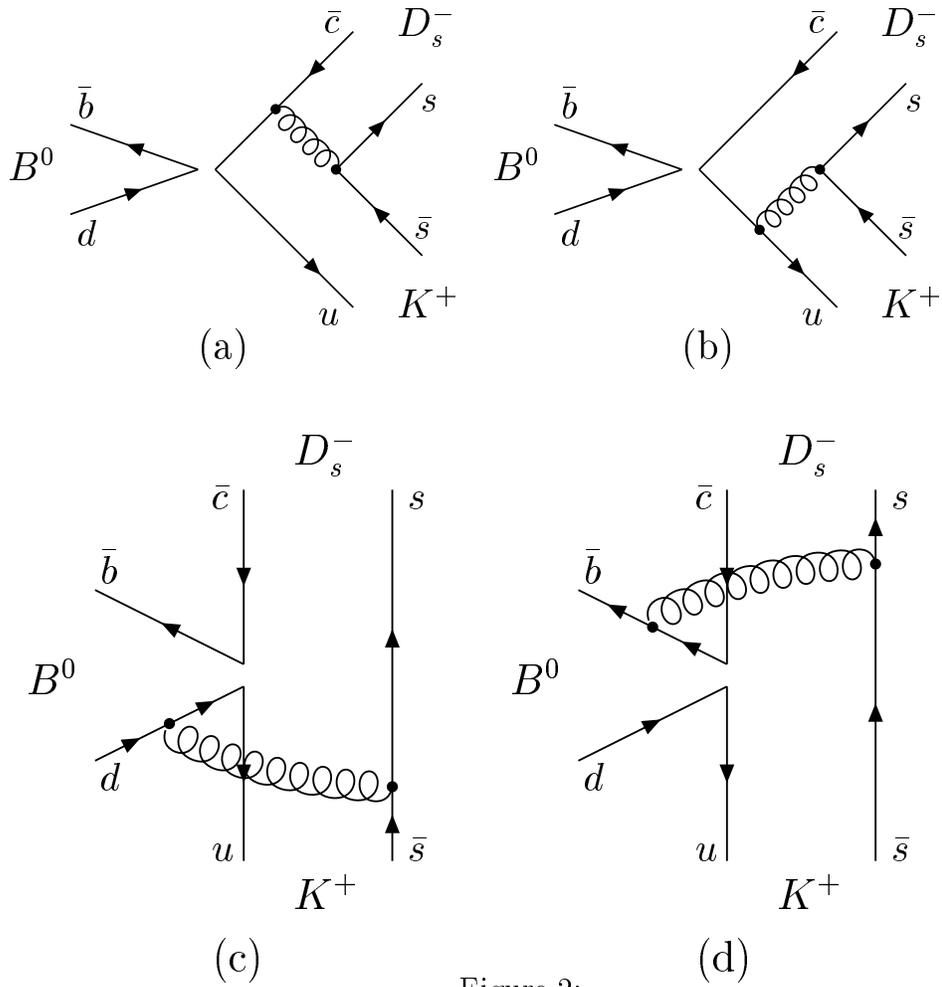
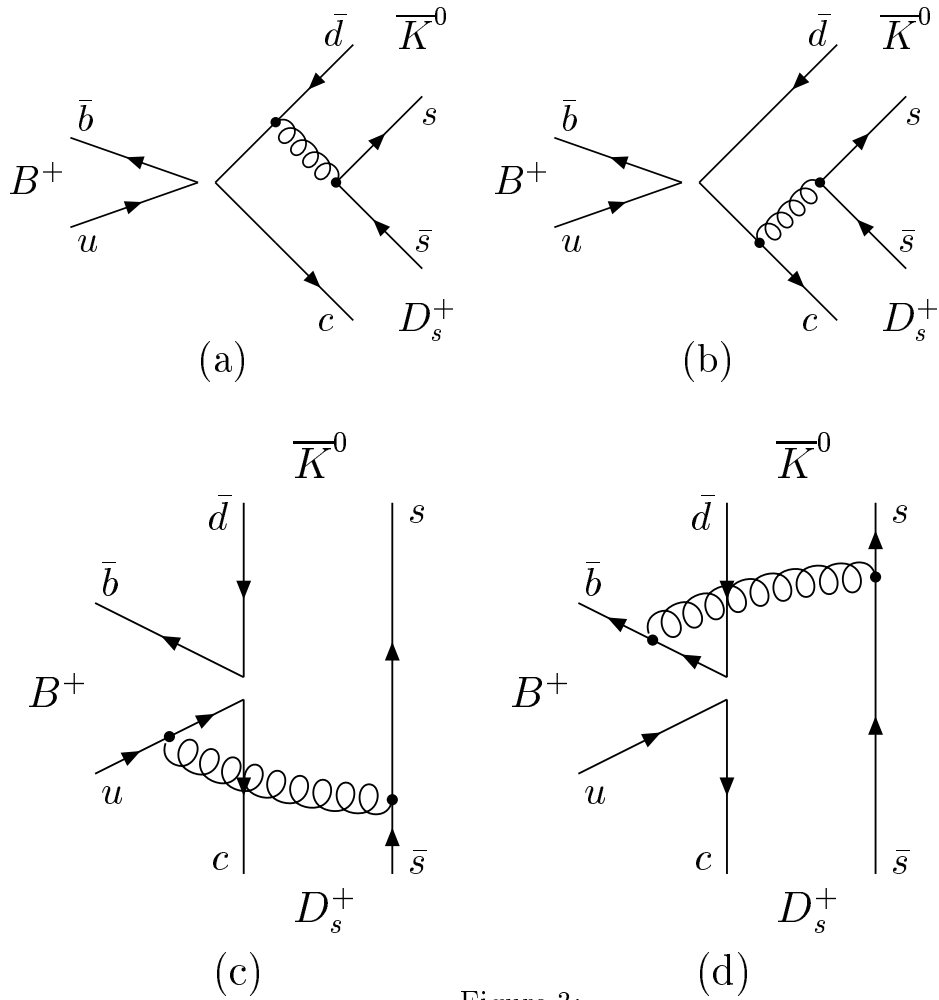
\end{document}